\documentstyle[12pt]{article}

\begin{document}

\begin{titlepage}

\title{Comment on ``Experimental demonstration 
of the violation of local realism without 
Bell inequalities'' by 
Torgerson et al.\footnote{Phys. Lett. A {\bf 214}, 
316-318 (1996).}}

\author{Ad\'{a}n Cabello\thanks{Present address: 
{\em Departamento de F\'{\i}sica Aplicada, Universidad de Sevilla, 
41012 Sevilla, Spain}. Electronic 
address: fite1z1@sis.ucm.es}\\
{\em Departamento de F\'{\i}sica Te\'{o}rica,}\\
{\em Universidad Complutense, 28040 Madrid, Spain.}\\
\\
Emilio Santos\\
{\em Departamento de F\'{\i}sica Moderna,}\\
{\em Universidad de Cantabria, 39005 Santander, Spain.}}

\date{\today}

\maketitle


\begin{abstract}
We exhibit a local-hidden-variable model in agreement 
with the results of the two-photon coincidence experiment 
made by Torgerson et al. [Phys. Lett. A 204 (1995) 323].
The existence of any such model shows that the experiment 
does not exclude local realism.

\end{abstract}

\end{titlepage}

In a recent paper \cite{TBMM}, Torgerson et al. 
(hereafter referred to as TBMM) claim that their two-photon 
coincidence experiment, based on the ideas of Hardy \cite{Hardy}, 
demonstrates the violation of local realism. In particular, they claim 
that even if three of the four measured probabilities are not 
exactly zero, as required for Hardy's argument, the results of 
the experiment still contradict local realism by about 45 standard 
deviations. We maintain that the experiment refutes only a restricted 
family of local hidden-variable (LHV) theories containing additional 
assumptions. To support this point of view, we present a LHV model 
in agreement with {\em actual} results of the TBMM experiment (Table 
1 in Ref. \cite{TBMM}). The mere existence of any such model 
(which could be in disagreement with some other measurements with the 
same experimental set up) proves that local realism is not refuted by 
the experiment \cite{TBMM}. Then we investigate what supplementary 
assumptions considered by TBMM are violated in our model.

It has already been proved \cite{Santos} that no experiment involving 
only coincidence detection rates may refute the whole family of LHV 
theories without supplementary assumptions; an argument more specific 
for the commented experiment follows. Table 1 in Ref. \cite{TBMM} is 
reproduced if the joint probability that, given one photon 
in each arm of the TBMM arrangement (see also Ref. \cite{TBM}), 
the photon in arm 1 is detected with the polarizer set to the angle 
$\theta _1$  and the photon in arm 2 is detected with 
the polarizer set to angle $\bar \theta _2=\theta _2+\frac{\pi}{2}$,
is of the form
\begin{equation}
P_{12}(\theta _1,\,\bar \theta _2) = N 
\left|\,{\cos\theta _1\cos \theta _2+{{\left| {R} 
\right|^{2}} \over {\left| {T} 
\right|^{2}}}\,\sin \theta _1\sin \theta _2}\,
\right|^{2}\,.
\label{uno}
\end{equation}
Any LHV model must express that joint probability as
\begin{equation}
P_{12}(\theta _1,\,\bar \theta _2)=\int {P_1(
\theta _1,\,\lambda )\,P_2(\bar \theta _2,\,\lambda )\,\rho(
\lambda )\,\mbox{d} \lambda }\,,
\label{dos}
\end{equation}
where $\lambda $ denotes collectively the hidden-variables, 
$\rho (\lambda )$ is the joint probability distribution of 
these hidden-variables (consequently, $\rho (\lambda )$ 
must satisfy both $\rho(\lambda )\ge 0$ and
$\int {\rho (\lambda )\,\mbox{d} \lambda }=1$), 
$P_1(\theta _1,\,\lambda )$  is the probability that 
the photon in arm 1 is detected with the polarizer 
set to the angle $\theta _1$, and analogously 
$P_2(\bar \theta _2,\, \lambda )$ (therefore 
$0\le P_1(\theta _1,\,\lambda ),\,\,P_2(\bar 
\theta _2,\,\lambda )\le 1$).

The model we propose has as hidden-variables 
two unitary vectors, $\mbox{\boldmath $u$}_1$ and $\mbox{\boldmath $u$}_2$.
We define
\begin{equation}
\rho (\lambda )\,\mbox{d} \lambda ={3 \over {(4\pi )^2}}\,
(\mbox{\boldmath $u$}_1\cdot {\cal R} \,\mbox{\boldmath $u$}_2)^2\,
\mbox{d} ^2\mbox{\boldmath $u$}_1\,\mbox{d} ^2\mbox{\boldmath $u$}_2\,,
\label{tres}
\end{equation}
where ${\cal R}$ denotes a rotation of angle $\varphi $ in the $x$-$y$ 
plane and
\begin{equation}
\cos \varphi ={{\left| {\,R\,} \right|^{2}} 
\mathord{\left/ {\vphantom {{\left| {\,R\,} \right|^{2}} 
{\left| {\,T\,} \right|^{2}}}} \right. 
\kern-\nulldelimiterspace} {\left| {\,T\,} 
\right|^{2}}}\,,
\label{cuatro}
\end{equation}
$R$ and $T$ being the reflectivity and transmissivity coefficients of 
the beam splitter. For the definition (\ref{cuatro}) we have assumed  
$\left| {\,R\,} \right|<\left| {\,T\,} \right|$, as in Ref.
\cite{TBMM}; if not, the changes $\theta _j\to \frac{\pi}{2}-\theta _j$
$(j=1,\,2)$, 
$\left| {\,R\,} \right|\leftrightarrow 
\left| {\,T\,} \right|$, would lead to an 
adequate model. We also define
\begin{equation}
P_1(\theta _1,\,\mbox{\boldmath $u$}_1)=C\kern 1pt \pi \kern 1pt 
\varepsilon \,
f(\mbox{\boldmath $u$}_1-\mbox{\boldmath $r$}_1)\,,
\label{cinco}
\end{equation}
\begin{equation}
\,P_2(\bar \theta _2,\,\mbox{\boldmath $u$}_2)=C\kern 1pt \pi 
\kern 1pt \varepsilon \,
f(\mbox{\boldmath $u$}_2-\mbox{\boldmath $r$}_2)\,,
\label{seis}
\end{equation}
where $0<C\le 1$, $0<\varepsilon <<1$,
\begin{equation}
f(\mbox{\boldmath $x$})=\left\{ {\matrix{\hfill {(\pi \kern 1pt \kern 1pt 
\varepsilon )^{-1}\quad \mbox{if}\quad \left| 
{\kern 1pt \mbox{\boldmath $x$}\kern 1pt } 
\right|^{2}\le \varepsilon }\cr
\hfill {0\;\quad \quad \mbox{if}\quad \left| 
{\kern 1pt \mbox{\boldmath $x$}\kern 1pt } 
\right|^{2}>\varepsilon }\cr
}} \right.\,,
\label{siete}
\end{equation}
and
\begin{equation}
\mbox{\boldmath $r$}_1=(\sin \theta _1,\,0,\,\cos \theta _1)\,,
\label{ocho}
\end{equation}
\begin{equation}
\mbox{\boldmath $r$}_2=(\sin \theta _2,\,0,\,\cos \theta _2)\,.
\label{nueve}
\end{equation}
With the above definitions and from expression (\ref{dos}), we obtain
\begin{equation}
P_{12}={3 \over {16}}\,C^2\kern 1pt \varepsilon ^2
(\mbox{\boldmath $r$}_1\cdot {\cal R} 
\, \mbox{\boldmath $r$}_2) ^2 + o(\varepsilon ^3)\,.
\label{diez}
\end{equation}
which is in agreement with (\ref{uno}) except for terms of order 
$\varepsilon ^3$.
The results of Table 1 in Ref. \cite{TBMM} can be reproduced if 
$\varepsilon$ is smaller than the error in the measurements.

Since the conditions required for Hardy's argument do not strictly 
occur in real experiments (see (14) in Ref. \cite{TBMM}), additional 
assumptions and a different argument are required in order to draw 
a conclusion from the experimental data. TBMM explicitly assume fair 
sampling. This implies in particular that (a) the photon losses in 
the polarizers and (b) the efficiencies of photodetectors behind the 
polarizers are both independent of the polarization of the photons. 
(b) can be circumvented since in Ref. \cite{TBMM} photodetectors 
receive 
photons with the same polarization. However, real polarizers do not 
satisfy (a). Neither does our model. The equality
\begin{equation}
P_{12}(\theta _1,\,\theta _2)+P_{12}(\theta _1,\,\bar \theta _2)=
P_{12}(\theta _1,\,\theta _{20})+P_{12}(\theta _1,\,\bar \theta _{20})\,,
\label{once}
\end{equation}
where $\theta _2$ and $\theta _{20}$ are two alternative settings 
for the polarizer 2, is 
fulfilled by the model only at the lowest order in 
$\varepsilon$. Violating 
(\ref{once}) means that photon absorption of polarizer 2 might depend 
on the orientation of the polarizer 2 (and similarly for polarizer 1). 
Since the model violates (\ref{once}), it also violates Eqs. 
(\ref{dos}) and (\ref{tres}) in Ref. \cite{TBMM}; but the argument 
of TBMM 
(see (15) and the following paragraphs in Ref. \cite{TBMM}) is 
essentially based on (\ref{dos}) and (\ref{tres}) and therefore cannot
be applied. 

In addition to our previous comments, we find a more profound 
criticism to Ref. \cite{TBMM}. The final step of TBMM's reasoning (in 
particular, their conclusion that the experimental data show a 
contradiction with local realism of about 45 standard deviations) 
is based on the following (mathematically incorrect) assumption
\begin{equation}
P_{12}(\theta _{10},\,\theta _{20})=P_{12}(\theta _1,\,\theta _2)\,
P_{12}(\theta _{20}\,|\,\theta _1)\,P_{12}(\theta _{10}\,|\,\theta _2)\,.
\label{doce}
\end{equation}
Under (\ref{doce}) lies a {\em na\"{\i}ve} idea generally attributed 
to LHV theories: if a photon is detected behind a polarizer oriented 
in the $\theta _1$, that is because the photon {\em had} a 
polarization $\theta _1$; but what 
a LHV theory actually says is that the photon has some $\lambda $ and 
that $P_1(\theta _1,\,\lambda )$, $P_1(\theta _{10},\,\lambda )$, etc.
exist. In LHV, the subset of $\lambda $ implicated in 
$P_{12}(\theta _1,\,\theta _2)$ might be 
different from the subset of $\lambda $ implicated in 
$P_{12}(\theta _1,\, - )$ and both might be 
different from the subset of $\lambda $ implicated in 
$P_{12}(- ,\,\theta _2)$. Therefore the 
equality (\ref{doce}) might not be fulfilled and no proof to refute 
local realism can be based on it.

The indubitable pedagogical value \cite{Mermin94a,Mermin94b} 
of the argument of Hardy \cite{Hardy}, or the possibility of 
implementing it in many different physical ({\em gedanken}) 
contexts \cite{Hardy,YS93,Freyberger}, does not mean that {\em actual} 
experiments based on Hardy's ideas will lead to more conclusive 
tests to exclude local realism than those based on Bell 
inequalities \cite{Bell}. This point is also stressed 
in Refs. \cite{Mermin94a,Mermin94b}. As Mermin perfectly
sums up in Ref. \cite{Mermin94a}, ``Hardy's four questions 
provide a rather weak basis for a laboratory violation of 
the experimentally relevant inequality'' 
(although ``they reign supreme in the {\em gedanken} realm'').\\

The authors would like to thank Guillermo Garc\'{\i}a Alcaine 
for discussions on this topic. One of us (E.S.)
acknowledges financial support from DGICYT 
Project PB-92-0507 (Spain).



\begin{thebibliography}{99}
\bibitem{TBMM} J.R. Torgerson, D. Branning, C.H. Monken 
and L. Mandel, Phys. Lett. A 204 (1995) 323.
\bibitem{Hardy} L. Hardy, Phys. Rev. Lett. 68 (1992) 2981; 
71 (1993) 1665; Phys. Lett. A 167 (1992) 17.
\bibitem{Santos} E. Santos, Phys. Rev. A 46 (1992) 3646.
\bibitem{TBM} J.R. Torgerson, D. Branning and L. Mandel, 
Appl. Phys. B 60 (1995) 267.
\bibitem{Mermin94a} N.D. Mermin, Phys. Today  
(June 1994) 9; (November 1994) 119.
\bibitem{Mermin94b} N.D. Mermin, Am. J. Phys. 62 (1994) 880.
\bibitem{YS93} B. Yurke and D. Stoler, 
Phys. Rev. A 47 (1993) 1704.
\bibitem{Freyberger} M. Freyberger, 
Phys. Rev. A 51 (1995) 3347.
\bibitem{Bell} J.S. Bell, Physics 1 (1964) 195.
\end{thebibliography}
\end{document}